\documentclass{article}

\usepackage[utf8]{inputenc}
\usepackage{amsmath}
\usepackage{geometry}
\usepackage{cite} % For bibliography management
\usepackage{hyperref} % For clickable links (optional, but good practice)

\hypersetup{
    colorlinks=true,
    linkcolor=blue,
    filecolor=magenta,      
    urlcolor=cyan,
    citecolor=green,
}

\title{Understanding the Algorithm Behind Audio Key Detection}
\author{Henrique Perez Gomes da Silva \\ henrique.perez@lumorobotics.com}
\date{May 22, 2025}

\begin{document}

\maketitle

\begin{abstract}
The determination of musical key is a fundamental aspect of music theory and perception, providing a harmonic context for melodies and chord progressions. Automating this process, known as automatic key detection, is a significant task in the field of Music Information Retrieval (MIR). This article outlines an algorithmic methodology for estimating the musical key of an audio recording by analyzing its tonal content through digital signal processing techniques and comparison with theoretical key profiles.
\end{abstract}

\section{Introduction}
Musical key provides a structural and perceptual framework for much of Western music. The ability to automatically identify the key of a piece of music from an audio recording has numerous applications, ranging from musicological analysis and recommendation systems to tools for DJs and musicians. The algorithm described herein leverages concepts from music cognition and signal processing to achieve this task. It involves extracting a representation of the audio's pitch class content and comparing it against standard templates for major and minor keys.

\section{Methodology}
The core of the described key detection algorithm involves representing musical keys as pitch class distributions, processing the input audio to extract its own pitch class distribution, and then finding the best match between the audio's characteristics and the predefined key representations.

\subsection{Representation of Musical Keys}
Musical keys are characterized by distinct patterns of emphasis on the twelve chromatic pitch classes: C, C\#, D, D\#, E, F, F\#, G, G\#, A, A\#, and B. These patterns can be represented as \textbf{key profiles}, which are 12-element vectors ($P$) where each element $P_k$ (for $k = 0, \dots, 11$) corresponds to the expected perceptual importance or stability of a pitch class within that key. For this algorithm, initial templates for C Major ($P_{Cmaj}$) and C Minor ($P_{Cmin}$) are defined based on established psychoacoustic models, such as those proposed by Krumhansl \cite{Krumhansl1990}. For instance, a C Major template might be represented numerically as $[6.35, 2.23, 3.48, 2.33, 4.38, 4.09, 2.52, 5.19, 2.39, 3.66, 2.29, 2.88]$.

To ensure that comparisons are based on the relative distribution of pitches rather than their absolute magnitudes, these profiles are \textbf{normalized}. This is achieved by dividing each profile vector $P$ by its Euclidean norm $\|P\|$, yielding a unit vector $P_{norm}$:
\begin{equation}
P_{norm} = \frac{P}{\|P\|}
\end{equation}
The Euclidean norm is calculated as:
\begin{equation}
\|P\| = \sqrt{\sum_{k=0}^{11} P_k^2}
\end{equation}
From these base C Major and C Minor profiles, a complete set of 24 key profiles (12 major and 12 minor) is generated. This is accomplished by \textbf{circularly shifting} the elements of the base profiles. If $P_{base}$ is a base profile (e.g., $P_{Cmaj}$), a new profile $P_{shifted}$ corresponding to a key $s$ semitones above the base key is generated such that its $k$-th element is:
\begin{equation}
P_{shifted}[k] = P_{base}[(k-s) \pmod{12}]
\end{equation}
This process is repeated for all 12 chromatic steps ($s=0, \dots, 11$) for both major and minor templates.

\subsection{Audio Signal Processing and Feature Extraction}
The first step in analyzing an audio recording is to \textbf{load the audio data}. This involves decoding the audio file format into a digital representation, specifically a one-dimensional array of discrete time samples $x[n]$ that represents the amplitude of the audio signal over time. Alongside the waveform, the \textbf{sampling rate} ($sr$), which is the number of samples per second, is also obtained.

A critical stage is the extraction of \textbf{chroma features} \cite{Fujishima1999, Bartsch2001}. This process begins by performing a \textbf{Short-Time Fourier Transform (STFT)} \cite{Rabiner1978, Muller2015}. The input audio waveform $x[n]$ is segmented into short, usually overlapping, frames. For each frame $m$, indexed by its starting sample $mH$ (where $H$ is the hop length in samples, e.g., 512 samples in the described implementation), a window function $w[n]$ of length $L$ is applied. The Discrete Fourier Transform (DFT) is then computed for each windowed frame. The $k$-th DFT coefficient for frame $m$, $X_m[k]$, is given by:
\begin{equation}
X_m[k] = \sum_{n=0}^{L-1} x[mH+n]w[n] e^{-j\frac{2\pi kn}{L}}
\end{equation}
where $L$ is the frame length (FFT window size, e.g., 4096 points in the described implementation) and $k$ is the frequency bin index.

The resulting magnitude spectrogram, $|X_m[k]|$, which is a time-frequency representation showing the energy at each frequency bin for each time frame, is then transformed into a \textbf{chromagram}. This transformation maps the spectral energy onto the 12 pitch classes. Conceptually, frequencies $f$ from the spectrum are first mapped to MIDI pitch numbers $M_{pitch}$:
\begin{equation}
M_{pitch} = 69 + 12 \log_2\left(\frac{f}{440 \text{ Hz}}\right)
\end{equation}
These MIDI pitches are then mapped to pitch classes $pc$ (integers from 0 to 11):
\begin{equation}
pc = M_{pitch} \pmod{12}
\end{equation}
The chroma feature $C_m(pc)$ for a given pitch class $pc$ in frame $m$ is calculated by summing the spectral magnitudes (or energies $|X_m[k]|^2$) of all frequency bins $k$ that fall into that pitch class across different octaves:
\begin{equation}
C_m(pc) = \sum_{k \in \text{Bins}(pc)} |X_m[k]|
\end{equation}
where $\text{Bins}(pc)$ represents the set of FFT frequency bin indices corresponding to the pitch class $pc$. This yields a 12-element chroma vector $C_m$ for each frame $m$. The sequence of these vectors forms the chromagram.

To obtain a single, time-invariant chroma vector ($\bar{C}$) that represents the overall tonal content of the audio piece, the chromagram is averaged across all $N_F$ time frames:
\begin{equation}
\bar{C}(pc) = \frac{1}{N_F} \sum_{m=0}^{N_F-1} C_m(pc)
\end{equation}
for each pitch class $pc$. If this mean chroma vector $\bar{C}$ is not a zero vector (indicating the audio is not silent), it is also \textbf{normalized} by dividing it by its Euclidean norm:
\begin{equation}
\bar{C}_{norm} = \frac{\bar{C}}{\|\bar{C}\|}
\end{equation}
This normalization is identical in form to Equation (1) and (2).

\subsection{Key Estimation via Profile Matching}
The final step in determining the musical key is to compare the normalized mean chroma vector derived from the audio signal, $\bar{C}_{norm}$, with each of the 24 pre-computed and normalized key profiles, $P_{j,norm}$, where $j$ indexes the 24 possible keys. This comparison quantifies the similarity between the audio's tonal content and the characteristic tonal patterns of each major and minor key.

The \textbf{correlation} between the audio's chroma vector and each key profile serves as this similarity measure. This is computed using the \textbf{dot product}. For two unit-normalized vectors, $\bar{C}_{norm}$ and $P_{j,norm}$, their dot product is:
\begin{equation}
corr_j = \bar{C}_{norm} \cdot P_{j,norm} = \sum_{k=0}^{11} (\bar{C}_{norm})_k \cdot (P_{j,norm})_k
\end{equation}
Since both vectors are normalized to unit length, this dot product is equivalent to the cosine of the angle $\theta_j$ between them, $corr_j = \cos(\theta_j)$. A higher dot product value, approaching 1, signifies a greater similarity.

The key profile $P_j$ that yields the maximum correlation score is identified as the estimated key of the audio:
\begin{equation}
Key_{estimated} = \text{argmax}_{j} (corr_j)
\end{equation}
The specific name of this key (e.g., "G Major", "D minor") is then reported. The magnitude of this highest correlation score can also provide an indication of the confidence in the key estimation.

Optionally, the detected standard key notation can be translated into alternative systems. For instance, the \textbf{Camelot Wheel} notation, popular among DJs for harmonic mixing, maps standard key names to a numerical and letter-based system. This is typically achieved using a predefined lookup table.

The robustness of the algorithm is enhanced by incorporating preliminary checks. For example, it verifies if the input audio file is of sufficient length to allow for meaningful STFT analysis. If an audio segment is predominantly silent ($\|\bar{C}\| \approx 0$), key estimation might be bypassed.

\section{Conclusion}
The described methodology provides a computational framework for estimating the musical key of an audio recording. By combining established signal processing techniques for feature extraction, such as the Short-Time Fourier Transform and chroma analysis, with a profile-matching approach rooted in music theory and psychoacoustics, this algorithm automates a crucial aspect of music analysis. Such automated key detection systems have valuable applications in music information retrieval, including organizing large music collections, generating music recommendations, and providing tools for musicians and DJs.

\end{document}